\documentclass[prl,superscriptaddress,unsortedaddress,twocolumn,%
  showpacs,preprintnumbers,amsmath,amssymb]{revtex4-1}
\usepackage{graphicx}
\usepackage[breaklinks,colorlinks=true]{hyperref}
\usepackage{slashed}

\newcommand{\instate}{|\text{in}\rangle}
\newcommand{\outstate}{|\text{out}\rangle}
\newcommand{\statein}{\langle\text{in}|}
\newcommand{\stateout}{\langle\text{out}|}
\newcommand{\Sprop}{S_{\text{in}}^{\text{c}}}
\newcommand{\Dprop}{\Delta_{\text{in}}^{\text{c}}}

\newcommand{\bB}{\boldsymbol{B}}
\newcommand{\bE}{\boldsymbol{E}}

\newcommand{\calO}{\mathcal{O}}

\newcommand{\llangle}{\langle\!\langle}
\newcommand{\rrangle}{\rangle\!\rangle}

\newcommand{\tr}{\mathrm{tr}\,}

\begin{document}
\title{Axial Ward identity and the Schwinger mechanism \\
  --- Applications to the real-time chiral magnetic effect and
  condensates ---}
\author{Patrick Copinger}
\author{Kenji Fukushima}
\author{Shi Pu}
\affiliation{Department of Physics, The University of Tokyo, %
             7-3-1 Hongo, Bunkyo-ku, Tokyo 113-0033, Japan}
\begin{abstract}
  We elucidate chirality production under parity breaking constant
  electromagnetic fields, with which we clarify qualitative
  differences in and out of equilibrium.  For a strong magnetic field
  the pair production from the Schwinger mechanism increments the
  chirality.  The pair production rate is exponentially suppressed
  with mass according to the Schwinger formula, while the mass
  dependence of chirality production in the axial Ward identity
  appears in the pesudo-scalar term.  We demonstrate that
  in equilibrium field theory calculus the axial anomaly is canceled
  by the pseudo-scalar condensate for any mass.  In a real-time
  formulation with in- and out-states, we show that the axial Ward
  identity leads to the chirality production rate consistent with the
  Schwinger formula.  We illuminate that such an in- and out-states
  formulation makes clear the chiral magnetic effect in and out of
  equilibrium, and we discuss further applications to real-time
  condensates.
\end{abstract}

% 25.75.-q Relativistic heavy-ion collisions
% 25.75.Nq Quark deconfinement, quark-gluon plasma production, and phase transitions
% 21.65.Qr Quark matter
% 12.38.-t Quantum chromodynamics
%\pacs{xxx}
\maketitle

%%%%%%%%%%   Introduction   %%%%%%%%%%
\paragraph*{Introduction:}
Chirality is a topical keyword for anomalous phenomena in
physics and related subjects.  In the context of high-energy physics in
which the fermion mass is often neglected, the chirality and the
helicity are identifiable, which has also motivated a modern
redefinition of chirality in chemistry~\cite{BARRON1986423}.

The most notable feature of chirality in relativistic fermionic
systems is the realization of the quantum anomaly.  Since relativistic
fermionic dispersion relations are realized in 2D and 3D materials, as
in the Weyl and Dirac
semimetals~\cite{PhysRevB.85.195320,PhysRevB.88.125427,Neupane2014,Liu2014},
it is of paramount importance to probe the chiral anomaly in
laboratory experiments, not only in quantum chromodynamics (QCD) but
also in optical environments.  One proposed signature for the chiral
anomaly is the negative magnetoresistance~\cite{Son:2012bg}, which
is a signal of chiral anomaly through the chiral magnetic
effect~\cite{Fukushima:2008xe}.  For the first experimental detection
as well as simplified theoretical arguments, see
Ref.~\cite{Li:2014bha}, and for the resummed field-theoretical
calculation of the negative magnetoresistance, see
Ref.~\cite{Fukushima:2017lvb}.

In all ideas to access the chiral anomaly, the generation of finite
chirality imbalance is indispensable.  The simplest optical setup is,
as discussed in Ref.~\cite{Fukushima:2010vw}, parallel electric and
magnetic fields.  Then, the chirality production rate is related to
the Schwinger mechanism as used in
Refs.~\cite{Fukushima:2010vw,Warringa:2012bq}, and at the same time it
is dictated by the axial Ward identity as argued in
Ref.~\cite{Iwazaki:2009bg}.  Such a simple electromagnetic
configuration is also useful to test ideas in the real-time numerical
simulations~\cite{Fukushima:2015tza,Mueller:2016ven}.

Even though the parallel electromagnetic fields are simple to treat, there
are still some controversies especially on different manifestations of
the chiral anomaly in and out of equilibrium.  In this Letter we
clarify these controversies by addressing the following two closely
related problems, namely:
\begin{itemize}
\item The effect of fermion mass $m$;  it is quite often assumed that
  the mass dependent term can be dropped from the axial Ward identity
  if $m=0$, but this is not always justified.
\item Equilibrium and real-time observables;  the $m$ dependence is
  totally different depending on how to take the expectation value in
  the presence of electric fields.
\end{itemize}
Answering these questions will naturally lead us to a clear picture of
chiral dynamics.  Moreover, we will see that our present
considerations have many applications to be studied in the future.
\vspace{0.5em}

\begin{figure}
  \includegraphics[width=0.4\columnwidth]{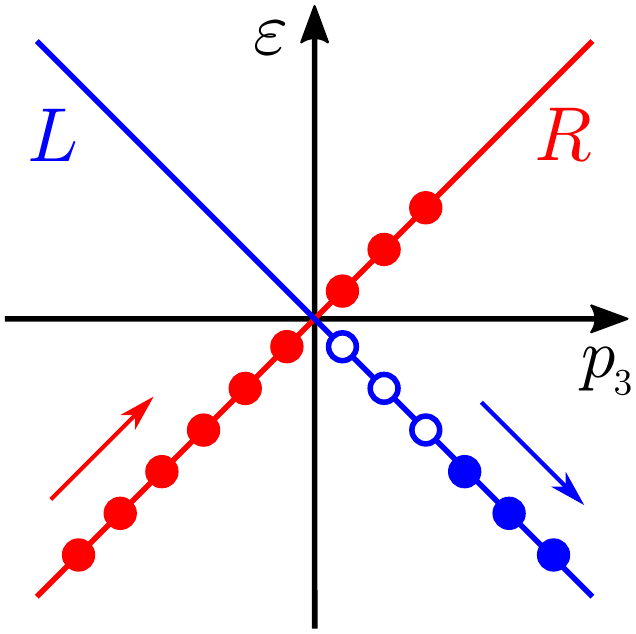}
  \caption{Schematic picture of the pair production in a reduced
    (1+1)-dimensional system in the LLLA; right-handed particles
    and left-handed antiparticles and thus net chirality are generated
    with parallel $E$ along the 3-axis.}
  \label{fig:dispersion}
\end{figure}

%%%%%%%%%%   An enigma   %%%%%%%%%%
\paragraph*{An enigma:}
We choose constant and parallel electric $E$ and magnetic $B$ fields
in the 3-axis direction.  Then, the celebrated formula for the
Schwinger mechanism reads,
\begin{equation}
  \omega = \frac{e^2 E B}{4\pi^2}
  \coth\biggl(\frac{B}{E}\pi\biggr)
  \exp\biggl(-\frac{\pi m^2}{eE}\biggr)
\end{equation}
for the pair production rate (for a comprehensive review, see
Ref.~\cite{Dunne:2004nc}).  In a particular limit of strong $B$ (i.e.,
$\sqrt{eB}$ being the largest mass scale in a system), the spin
direction is completely aligned along $B$, so that particles have
definite chirality in a reduced (1+1)-dimensional system emerging in
the lowest Landau level approximation (LLLA).  The right-handed ($R$)
particles increase and the left-handed ($L$) particles decrease
creating $L$ antiparticles under $E$, as sketched in
Fig.~\ref{fig:dispersion}.

%---   figure   ---%

A pair of $R$ and $\bar{L}$ production thus changes the chirality by
two, leading to a relation as used in Ref.~\cite{Fukushima:2010vw},
\begin{equation}
  \omega\; \overset{B\gg E}{\longrightarrow}\;\;
  \frac{e^2 E B}{4\pi^2} \exp\biggl(-\frac{\pi m^2}{e E}\biggr)
  = \frac{1}{2}\partial_t n_5\,,
  \label{eq:schwinger}
\end{equation}
where $n_5$ is the chiral charge density, that is an expectation value
of $j_5^0$.

The right-hand side, $\partial_t n_5$, is dictated by the axial Ward
identity, i.e.,
$\partial_\mu j_5^\mu = -\frac{e^2}{16\pi^2}\epsilon^{\mu\nu\alpha\beta}
  F_{\mu\nu}F_{\alpha\beta} + 2m\bar{\psi} i \gamma_5\psi$
on the operator level, where
$\epsilon^{\mu\nu\alpha\beta}F_{\mu\nu}F_{\alpha\beta}=-8E B$ for
parallel $E$ and $B$ in the present setup.  After taking the
expectation value with $\langle j_5^i\rangle=0$ presumed, the
chirality production should follow,
\begin{equation}
  \partial_t \bar{n}_5 = \frac{e^2 EB}{2\pi^2}
  + 2m \langle\bar{\psi}i\gamma_5 \psi\rangle\,.
  \label{eq:ward}
\end{equation}
(The reason for changing $n_5$ by $\bar{n}_5$ will be explained below.)
One might have thought that Eqs.~\eqref{eq:schwinger} and
\eqref{eq:ward} are consistent if $m=0$, which is frequently assumed
in the literature.  The justification is not so trivial, however,
even for the $m=0$ case.  Because parallel $E$ and $B$ make a parity
breaking combination, $\langle\bar{\psi}i\gamma_5\psi\rangle \propto E
B$ is anticipated.  In fact, for the pseudo-scalar condensate,
Schwinger performed the calculation using the proper-time method to
discuss the interaction between the neutral meson and the proton,
which results in~\cite{Schwinger:1951nm}
\begin{equation}
  \bar{P}:= \langle\bar{\psi}i\gamma_5\psi\rangle
  = -\frac{e^2 EB}{4\pi^2 m}\,.
  \label{eq:pseudoscalar}
\end{equation}
This makes $\partial_t \bar{n}_5 = 0$ for any $m$ including even the
$m=0$ limit!  This apparent contradiction between $\omega$ and
$\partial_t \bar{n}_5$ is quite often overlooked in a na\"{i}ve
treatment of dropping $m\langle\bar{\psi}i\gamma_5\psi\rangle$ for
$m=0$.  As a matter of fact, it is well known that an $m=0$ Abelian
gauge theory could be quite different from a theory in the $m\to 0$
limit, where in the former the electric charge is completely
shielded~\cite{Fomin:1976am}.  In QCD, $m\approx 0$ only approximately
holds, so we should consider the latter limit judiciously.
\vspace{0.5em}

%%%%%%%%%%   In-state and out-state   %%%%%%%%%%
\paragraph*{In-state and out-state:}

To resolve this puzzle, the crucial observation is that the vacua at
$t\to\pm\infty$ are not identical when an $E$ field is imposed, even
if $E$ itself is static.  If we
take $A_0=0$ gauge, $A_3(t)$ is needed for $E$ along the 3-axis, and
the in-state $\instate$ and the out-state $\outstate$ are different by
$A_3(\pm\infty)$, which are connected by the Bogoliubov
transformation.  Let us introduce the following notation for the in-
and out-states expectation values;
\begin{equation}
  \langle\calO\rangle := \stateout \calO(t) \instate \,,\qquad
  \llangle\calO\rrangle := \statein \calO(t) \instate
  \label{eq:inout_def}
\end{equation}
for an operator $\calO(t)$ in the Heisenberg representation.  In
quantum field theory calculus the generating functional represents an
amplitude $\langle\text{out}|\text{in}\rangle$, and we must clearly
distinguish,
\begin{equation}
  \bar{n}_5 := \langle j_5^0 \rangle\,,\qquad
  n_5 := \llangle j_5^0 \rrangle\,.
\end{equation}
In principle, one can utilize the Schwinger
closed time path formalism to deal with $\llangle\calO\rrangle$.  The
straightforward approach is, however, technically complicated for our
present problem especially with $A_3(t)$.

\begin{figure}
  \includegraphics[width=\columnwidth]{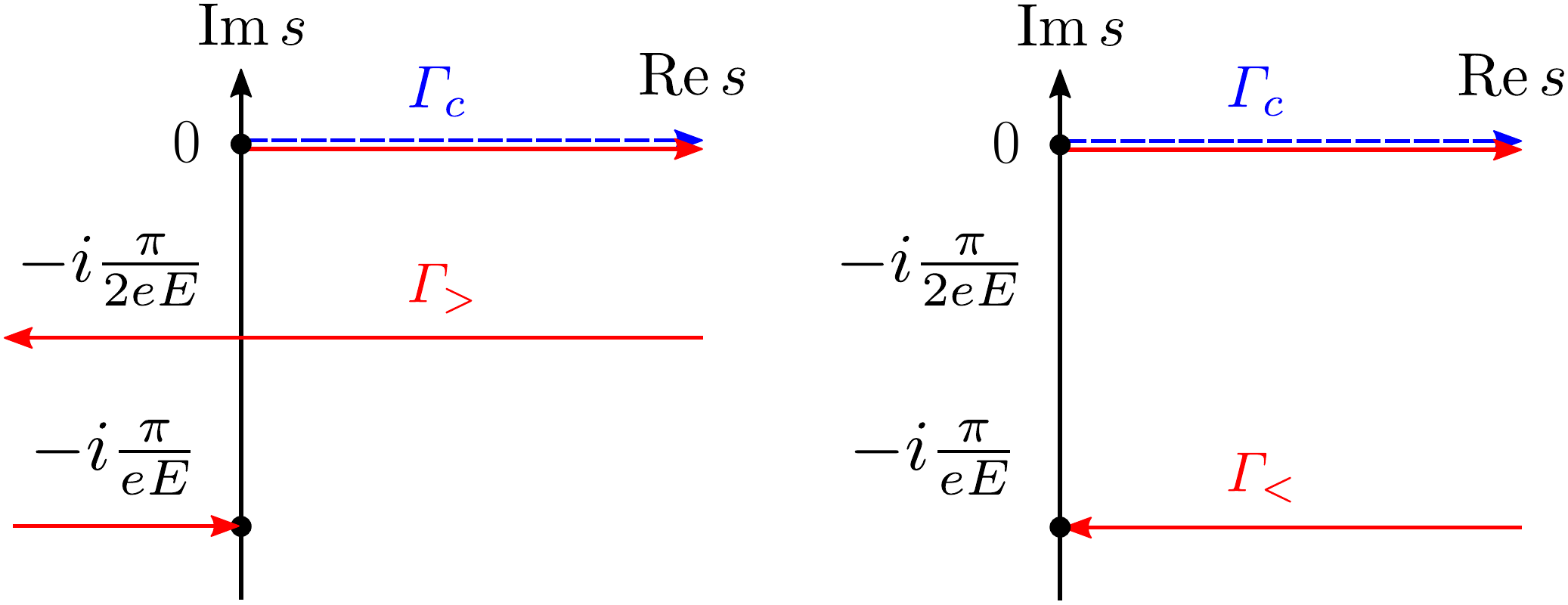}
  \caption{For the standard propagator, $i\langle
    T\psi(x)\bar{\psi}(y)\rangle$, the proper-time integration should
    go along $\Gamma_c$ (dashed line), while for $\Sprop$ the contour
    is complexified as $\Gamma_>$ for $z_3>0$ and $\Gamma_<$ for
    $z_3<0$.}
  \label{fig:contour}
\end{figure}

Fortunately, for the constant $E$ case, a much simpler formulation has
been known.  According to
Ref.~\cite{fradkin1991quantum} the two-point correlation function
takes the following proper-time representation,
\begin{equation}
  \Sprop(x,y) := i\llangle T\psi(x)\bar{\psi}(y)\rrangle
  = (i\slashed{D}_x + m)\Dprop(x,y)\,,
  \label{eq:Sin}
\end{equation}
where we can express $\Dprop(x,y)$ using $z=x-y$ and the proper-time
$s$ as
\begin{equation}
  \Dprop(x,y) = \biggl[ \theta(z_3)\int_{\Gamma^>} \!ds
    + \theta(-z_3)\int_{\Gamma^<} \!ds \biggr] g(x,y,s)\,.
\end{equation}
Here, the integration contours, $\Gamma^>$ and $\Gamma^<$, are shown
in Fig.~\ref{fig:contour}, respectively, where we introduced a
slightly different (but equivalent) representation from
Ref.~\cite{fradkin1991quantum}.

%---   figure   ---%
%---   figure   ---%

After some calculations we find the integration kernel with parallel
$E$ and $B$ as
\begin{equation}
  \begin{split}
  & g(x,y,s) =  \frac{e^2 EB}{(4\pi)^2} \sinh^{-1}(eEs)\sin^{-1}(eBs) \\
  &\qquad \times \exp\biggl[
    -i\Bigl(\frac{1}{2}eF\cdot \sigma + m^2\Bigr)s
      + i\varphi(x,y,s)\biggr]\,,
  \end{split}
\end{equation}
where $m^2$ should be understood as $m^2-i\epsilon$ for convergence in
Minkowskian spacetime.  We used a shorthand notation,
$F\cdot\sigma=F_{\mu\nu}\sigma^{\mu\nu}$, whose explicit form is
\begingroup
\setlength\arraycolsep{-5pt}
\begin{equation}
  e^{-i\frac{1}{2}eF\cdot \sigma s}
  = \begin{pmatrix}
    e^{-e(E-iB)s} & & & 0\\
     & e^{e(E-iB)s} & & \\
     & & e^{e(E+iB)s} & \\
    0 & & & e^{-e(E+iB)s}
    \end{pmatrix}
\end{equation}
\endgroup
in the chiral representation of $\gamma^\mu$.  The coordinate
dependent phase is
\begin{equation}
  \varphi(x,y,s) = \frac{1}{2}x\cdot eF\cdot y
  -\frac{1}{4}z\cdot \coth(eFs)\cdot eF\cdot z\,,
\end{equation}
where $x\cdot eF\cdot y=x_\mu eF^{\mu\nu} y_\nu$, etc.  Armed with the
explicit form of the two-point correlation function, we are ready for
concrete calculations.
\vspace{0.5em}

%%%%%%%%%%   An \'{e}claircissement   %%%%%%%%%%
\paragraph*{An \'{e}claircissement:}
Let us first consider,
\begin{equation}
  P := \llangle\bar{\psi}i\gamma_5\psi\rrangle
  = -\lim_{y\to x}\tr[\gamma^5 \Sprop(x,y)]\,.
\end{equation}
For this expectation value a term involving $\slashed{D}$ in
Eq.~\eqref{eq:Sin} vanishes in the $y\to x$ limit, and so
$P=-im\tr[\gamma^5 \Dprop(x,y\to x)]$, leading to
\begin{equation}
  P = -4i\frac{m e^2 EB}{(4\pi)^2} \int_{\Gamma} ds\,
    e^{-im^2 s} 
    = -\frac{e^2 EB}{4\pi^2 m} \Bigl[
      1 - e^{-\pi m^2/(eE)} \Bigr]\,,
  \label{eq:Pin}
\end{equation}
where $\Gamma$ is either $\Gamma^>$ or $\Gamma^<$, which is irrelevant
for the computation of $P$ since there is no pole in the integrand.
Then, with Eqs.~\eqref{eq:ward} and \eqref{eq:Pin}, we see that
Eq.~\eqref{eq:schwinger} holds as it should!  We note that
Eq.~\eqref{eq:Pin} was conjectured in Ref.~\cite{Warringa:2012bq}
from the Schwinger formula, but the clear recognition of in- and
out-states as revealed in this work was missing.

Interestingly, we can directly evaluate $n_5$ from Eq.~\eqref{eq:Sin}
to reach a consistent conclusion, i.e., $n_5=2\omega\cdot t$ correctly
(where $t$ appears from the phase space
volume~\cite{fradkin1991quantum}), while we find $\bar{n}_5=0$ as is
again consistent with $\partial_t \bar{n}_5=0$.  Although the details
for the $n_5$ and $\bar{n}_5$ calculations are extremely intriguing on
their own, we will spell out step-by-step procedures in a follow-up
publication.

Now, we have clarified that $n_5$ in the Schwinger mechanism should be
a real-time observable as defined by $\llangle\calO\rrangle$ in
Eq.~\eqref{eq:inout_def}, which has demystified consistency with the
axial Ward identity.  Then, it is interesting to consider what
$\partial_t \bar{n}_5=0$ means.  To answer this question it would be
convenient to Wick-rotate the time (not the proper-time) as $t\to\tau$
to switch to Euclidean theory, in which $\instate$ and $\outstate$ are
the ground states at $\tau=0$ and $\tau=\beta$, respectively.  Then,
what we computed by $\langle\calO\rangle$ corresponds to the
equilibrium expectation value at $\beta=\infty$ (i.e., zero
temperature).

The situation is even more transparent if we performed the Euclidean
Monte-Carlo calculation on the lattice.  As discussed in
Ref.~\cite{Yamamoto:2012bd}, even with Minkowskian $E$ (for which two
flavors are oppositely charged to avoid the sign problem, which may
well be called an ``isospin'' $E$), the lattice simulation always
measures $\langle\calO\rangle$ and cannot describe the real-time
particle production; only the charge distribution in the equilibrated
final state is observed.

We note that $\langle\partial_\tau\, j_5^\tau\rangle = 0$ resulting
from $\partial_t\bar{n}_5=0$ is what we must have in the $m=0$ limit.
The topological properties of the ground state are
characterized by the $\theta$ angle, and the ground state energy
becomes independent of $\theta$ if there is any $m=0$ fermion, which
constrains that the theory has only zero topological charge.  Then,
naturally, there is no nonzero topological charge and no chirality
flip.  In this way, $\partial_t \bar{n}_5=0$ is actually demanded for
$m=0$ in equilibrium and the careful $m\to0$ limit of the
second term in Eq.~\eqref{eq:ward}, which may not always vanish, is
crucial.  We must emphasize that it is astonishing that
$\partial_t \bar{n}_5=0$ holds for \textit{any} $m$, about which we
have no simple explanation.

One important extension along these lines of $n_5$ is found in the
calculation of the chirality fluctuation.  In particular, in the
high-energy nuclear collision experiments, fluctuations averaged over
many collision events or even with a single event are important
physical observables.  Then, it is an urgent task in theory to
compute,
$\chi_5 := (\llangle N_5^2 \rrangle - \llangle N_5 \rrangle^2)/V$
where $N_5:=\int d^3 x\, j_5^0(x)$ to quantify the idea of the local
parity violation.  Our setup with parallel $E$ and $B$ is very simple
but mimics the initial condition of the nuclear collision known as the
Glasma flux tubes.  Again, if we performed the lattice Monte-Carlo
simulation, we would get
$\bar{\chi}_5:=(\langle N_5^2\rangle - \langle N_5\rangle^2)/V$, which
is qualitatively different from our interested $\chi_5$.  As we
already saw, the disconnected piece,
$\llangle N_5\rrangle = \int d^3x\, 2\omega\cdot t$, grows linearly
with time, and the term $\propto t^2$ in $\llangle N_5^2\rrangle$ is
subtracted in $\chi_5$.  This also implies that a term $\propto t$
still remains in $\chi_5$, which is absent in $\bar{\chi}_5$.  We
actually find such a term $\propto t\cdot EB\,e^{-2\pi m^2/(eE)}$, all the
details about which shall be reported in a follow-up publication.
Instead, below, we shall focus on local observables involving only one
$\Sprop(x,y)$, namely, the expectation values of the current and the
scalar operators.
\vspace{0.5em}

%%%%%%%%%%   Chiral magnetic effect in and out of equilibrium   %%%%%%%%%%
\paragraph*{Chiral magnetic effect in and out of equilibrium:}
It is a straightforward exercise to compute the vector current
associated with the chiral magnetic effect.  In
Ref.~\cite{Fukushima:2010vw} the answer was inferred from the Lorentz
transformation of the Schwinger formula.  Although it
is a tedious calculation, particularly complex poles in the
proper-time integration need careful treatments, after all we arrive
at
\begin{equation}
  j^3 = \llangle \bar{\psi}\gamma^3\psi\rrangle
  = -\lim_{y\to x} \tr\bigl[\gamma^3 \Sprop(x,y)\bigr]
  = 2\omega\cdot t\,,
\label{eq:cme_out}
\end{equation}
which is the right answer for strong $B$;  generated $n_5$ is
immediately converted to $j^3$.  We note that the above result does
not rely on the LLLA;  we simply used the LLLA argument to relate the
axial Ward identity and the Schwinger formula, but our results of
$P$, $n_5$, $j^3$ are all valid beyond the LLLA{}.

Here, interestingly, a simple calculation yields
\begin{equation}
  \bar{j}^3 := \langle \bar{\psi}\gamma^3 \psi\rangle = 0
\label{eq:cme_in}
\end{equation}
in the same way to obtain $\bar{n}_5=0$.  This is a very important
result to demonstrate unambiguously that the chiral magnetic effect
does \textit{not} exist in equilibrium.  In short, in equilibrium,
there is no chirality generation, and there is no topological
current.

Such a statement about equilibrium chiral magnetic effect itself is
not quite new (see
Refs.~\cite{PhysRevLett.111.027201,Yamamoto:2015fxa}).  Some confusion
is attributed to a chiral chemical potential $\mu_5$;  with
$\mu_5$ the equilibrium chiral magnetic effect seemed to
exist~\cite{Fukushima:2008xe} but one should be aware of the fact that
the introduction of $\mu_5$ implicitly assumes a system out of
equilibrium (otherwise, $\mu_5$ should be vanishing).  Thus, $\mu_5$
was a very useful bookkeeping device to access the correct physics,
but it smeared qualitative differences in the chiral magnetic effect
in and out of equilibrium.  Equations~\eqref{eq:cme_out} and
\eqref{eq:cme_in} clearly show that the chiral magnetic effect is an
intrinsically real-time phenomenon.  Also we note that, if one
performed the lattice Monte-Carlo simulation with (Euclidean or
isospin Minkowskian) $E$ and $B$ trying to quantify a chiral magnetic
current, Eq.~\eqref{eq:cme_in} predicts that the lattice result for
the current should be zero.
Now, let us turn to another problem, that is, a scalar condensate.
Since we already discussed the pseudo-scalar condensates, $\bar{P}$
and $P$, it would be a quite natural extension of the study.
\vspace{0.5em}

%%%%%%%%%%   Dynamical chiral condensate   %%%%%%%%%%
\paragraph*{Dynamical chiral condensate:}
As an application of the insight we gained, let us consider the
expectation value of the scalar condensates (which are often called
the chiral condensates in QCD).  The scalar condensates are as
important as $\bar{P}$ and $P$ since they would induce constituent
masses.  In the absence of $E$, it is established that strong $B$ and
a tiny interaction would inevitably lead to a finite chiral
condensate, and this is commonly referred to as the magnetic
catalysis~\cite{Gusynin:1994re,Gusynin:1994xp}.  In this case the
induced chiral condensate (apart from interaction contributions)
reads,
\begin{align}
  \bar{\Sigma}\bigr|_{E=0} &:= \langle\bar{\psi}\psi\rangle_{E=0}
  = -\frac{eB}{4\pi^2}\, m\int_{\Gamma_c^{\Lambda}}
  \!\frac{ds}{s} e^{-i m^2 s} \cot(eB s) \notag\\
  & \simeq -\frac{eB}{4\pi^2}\, m\, \Gamma[0,m^2/\Lambda^2]\,,
  \label{eq:catalysis}
\end{align}
where we can imaginary-rotate~\footnote{To justify the imaginary
  rotation with singularities from $\coth(eBs)$ we should consider a
  nonzero $z_1$ and $z_2$ first, for which contours near and below
  these poles can be added without changing the integral.  Then the
  imaginary rotation does not hit any poles, and finally $z_1, z_2\to
  0$ can be taken.}
the proper-time $is\to s$ to make the above look identical to the
expression in Refs.~\cite{Gusynin:1994re,Gusynin:1994xp}.  We can also
approximate the incomplete gamma function as
$\Gamma[0,x]\sim -\gamma_{\rm E}-\ln x$ for small $x$, with
$\gamma_{\rm E}$ being the Euler-Mascheroni constant.  The ultraviolet
divergence is regularized by a shift of $\Gamma_c$ near the
origin as $s=0\to 1/\Lambda^2$, which is denoted by
$\Gamma_c^\Lambda$.  From the first to
the second line, $\coth(eBs)\sim 1$, corresponding to the LLLA, is
used.  The logarithmic singularity with respect to $m^2$ above is
translated into a term $\sim m^2\ln(\Lambda^2/m^2)$ in the effective
potential, which indicates a negative infinite curvature near
$m\sim 0$, and this gives rise to the magnetic catalysis.  In our
present formulation it is easy to include a finite $E$;
\begin{align}
  \bar{\Sigma} &= -\frac{e^2 EB}{4\pi^2}\, m
  \int_{\Gamma_c^\Lambda}
  ds\, e^{-i m^2 s} \cot(eBs)\coth(eEs) \notag\\
  &\simeq -\frac{eB}{4\pi^2}\, %e^{-m^2/\Lambda^2}
  m \biggl[\ln\frac{\Lambda^2\, e^{-\gamma_{\rm E}}}{2eE}
    - \mathrm{Re}\psi\Bigl(\frac{i m^2}{2eE}\Bigr)
    - \frac{i\pi}{e^{\pi m^2/(eE)}-1}\biggr]\,.
\label{eq:condensate_in}
\end{align}
Here, we used approximations, $\coth(eBs)\sim 1$ and
$e^{-m^2/\Lambda^2}\sim 1$, and wrote only terms nonvanishing in the
large-$\Lambda^2$ limit.  In the second line $\psi(x)$ represents
the digamma function.  Using an asymptotic expansion,
$\psi(x)\sim \ln x - 1/2x$ for large $x$, we can exactly recover
Eq.~\eqref{eq:catalysis} from the $eE\to 0$ limit of
Eq.~\eqref{eq:condensate_in}.

From Eq.~\eqref{eq:condensate_in} we see that the magnetic catalysis
is overridden by $eE$ and there is no longer a logarithmic singularity
with respect to $m^2$ (for related work on the phase structure with
parallel $E$ and $B$, see Ref.~\cite{Wang:2018gmj}).  Another
noticeable feature of Eq.~\eqref{eq:condensate_in} is that the scalar
condensate takes a complex value, which is analogous to a complex
chiral condensate at finite strong-$\theta$
angle~\cite{Boer:2008ct,Mameda:2014cxa}.  Since $\theta$ and
$\bE\cdot\bB$ share the same quantum number, we can naturally
anticipate the same behavior on the scalar condensate.  An interesting
point in Eq.~\eqref{eq:condensate_in} is that
$\mathrm{Im}\bar{\Sigma}$ takes a form of the Fermi-Dirac distribution
function with the energy over the temperature replaced by
$\pi m^2/(eE)$.
  
In view of the fact that $P$ turns out to be quite different from
$\bar{P}$ as in Eqs.~\eqref{eq:pseudoscalar} and \eqref{eq:Pin},
we may well anticipate such changes for the equilibrium and the
real-time or dynamical chiral condensates.  In the same way as $P$, we
can perform the calculation using $\Sprop(x,y)$, and then the
imaginary rotated ($is\to s$) expression takes the following form,
\begin{align}
  \Sigma &:= \llangle\bar{\psi}\psi\rrangle = -m\,\tr\bigl[
    \Dprop(x,y\to x)\bigr] \notag\\
  &= -\frac{e^2 EB}{4\pi^2}\, m\int_{1/\Lambda^2}^{\pi/eE-1/\Lambda^2}
  \!\!\!\!\!\!\!\! ds\, e^{-m^2 s} \coth(eBs)\cot(eEs) \notag\\
  &\simeq \Bigl[ 1 - e^{-\pi m^2/(eE)} \Bigr] \,\text{Re}\,
  \bar{\Sigma}\,,
  \label{eq:condensate_out}
\end{align}
where we dropped a term involving $\slashed{D}$ in the $y\to x$
limit, and kept only terms nonvanishing in the large-$\Lambda^2$
limit.  It should be noted that the ultraviolet divergences appear
from both edges of the integration range;  the integration without the
cutoff would diverge in the $z\to 0$ limit in exactly the same way
near $s=0$ and $\pi/(eE)$ (in the LLLA).  We thus regularized the
integration by shifting both edges equally by $1/\Lambda^2$.
Then, this overall factor in Eq.~\eqref{eq:condensate_out} happens to
be the same as that between $P$ and $\bar{P}$.  We remark that no
imaginary part appears in this case due to deformation of the
integration contour.

From the point of view of the spontaneous symmetry breaking,
Eq.~\eqref{eq:condensate_out} means that the
generation of the chiral condensate and thus the constituent mass is
further diminished as compared to the equilibrium case.  From the
dynamical point of view, Eq.~\eqref{eq:condensate_out} is a further
significant result.  It is sometimes a puzzling question whether the
Schwinger mechanism produces the bare particles or the dressed
particles.  In other words, the question is whether the Schwinger
critical mass is characterized by the bare mass or the constituent
mass.  Equation~\eqref{eq:condensate_out} indicates that, once the
pair production is activated with $e^{-\pi m^2/(eE)}\sim 1$, the dynamical
chiral condensate $\Sigma$ should melt, so that there is no
constituent mass (which is defined by a mean-field approximation for
the in-in expectation values).
\vspace{0.5em}

%%%%%%%%%%   Summary   %%%%%%%%%%
\paragraph*{Summary:}
We have clarified important differences associated with the in- and
out-states in the presence of electric field $E$.  In particular the
mass dependence appears quite different, which resolves some
controversies in the interpretations of the axial Ward identity, the
chiral magnetic effect, and the chiral condensate.  Here, to make our
point not blurred by technicalities, we limited ourselves to the
calculations involving one propagator only, namely, the expectation
values of the current and the scalar operators.
We will elsewhere discuss systematic computations of
higher order observables, such as the real-time chirality
fluctuations.
\vspace{0.5em}

\begin{acknowledgments}
The authors thank
Stefan~Fl\"{o}rchinger,
Xu-Guang~Huang, 
Niklas~Mueller, and
Naoto~Tanji for useful
comments and discussions.
K.F.\ is grateful for a warm hospitality at the Fudan University where
he stays as a Fudan University Fellow.
This work was supported by Japan Society for the Promotion of Science
(JSPS) KAKENHI Grant No.\ 18H01211.
S.P.\ was supported by JSPS post-doctoral fellowship for foreign
researchers.
\end{acknowledgments}

\bibliography{ward}
\bibliographystyle{apsrev4-1}

\end{document}